\documentclass[12pt]{article}
\usepackage{graphicx}
\usepackage{lscape}
\usepackage{citesort}

\begin{document}

\def\ds{\displaystyle}
\def\beq{\begin{equation}}
\def\eeq{\end{equation}}
\def\bea{\begin{eqnarray}}
\def\eea{\end{eqnarray}}
\def\beeq{\begin{eqnarray}}
\def\eeeq{\end{eqnarray}}
\def\ve{\vert}
\def\vel{\left|}
\def\ver{\right|}
\def\nnb{\nonumber}
\def\ga{\left(}
\def\dr{\right)}
\def\aga{\left\{}
\def\adr{\right\}}
\def\lla{\left<}
\def\rra{\right>}
\def\rar{\rightarrow}
\def\nnb{\nonumber}

\def\simlt{\stackrel{<}{{}_\sim}}
\def\simgt{\stackrel{>}{{}_\sim}}


\title{ {\Large {\bf Analysis of  the $B_q\to D_q(D_q^*) P$ and $B_q\to D_q(D_q^*) V$ decays within  the factorization approach in QCD} } }

\author{\vspace{1cm}\\
{\small K. Azizi$^1$ \thanks {e-mail: kazizi@dogus.edu.tr}~\,}, {\small R. Khosravi$^2$ \thanks {e-mail: khosravi.reza @ gmail.com}~\,, \small F. Falahati$^2$ \thanks {e-mail: phy1g832889 @shiraz.ac.ir}~\,} \\
{\small $^1$ Physics Division,  Faculty of Arts and Sciences, Do\u gu\c s University,}\\
{\small Ac{\i}badem-Kad{\i}k\"oy,  34722 Istanbul, Turkey}\\
 {\small $^2$ Physics Department , Shiraz University, Shiraz 71454,
Iran}\\}
\date{}

\begin{titlepage}
\maketitle \thispagestyle{empty}

\begin{abstract}
Using the  factorization approach and considering the contributions of the  current-current, QCD penguin and electroweak penguin operators at the leading approximation, the decay amplitudes and decay
widths of $B_q\to D_q(D_q^*) P$ and $B_q\to D_q(D_q^*) V$
transitions, where $q=u,d,s$ and P and V are pseudoscalar and vector
mesons, are calculated in terms of the transition form factors of the $B_q\to D_{q}$ and  $B_q\to D^{*}_{q}$. Having
computed those form factors in three-point QCD sum rules, the
branching fraction for these decays are also evaluated. A comparison
of our results with the predictions of the perturbative QCD as well as
the existing experimental data is presented.

\end{abstract}


\end{titlepage}

\section{Introduction}

With the chances that a lot of $B_{q}$ mesons will be produced in B
factories \cite{Aubert,Mori}, it would be possible to check the
two-body non-leptonic charmed decay modes $B_q\to D_q(D_q^*) P$ and
$B_q\to D_q(D_q^*) V$. Analyzing of such type decays could give
valuable  information about the origin of the CP violation, hadronic
flavor changing neutral currents, test of the standard model (SM),
constraints on new physic parameters as well as strong interactions
among the participating particles which provides valuable tests of
the QCD  factorization framework.

Theoretically, analyzing of the two-body B-decay amplitudes have
been started using the framework of  so called "naive factorization"
\cite{Bauer,Neubert,Ali1,Ali2,Chen}. This method for some decay channels is replaced by QCD
factorization \cite{Beneke1,Beneke2} since it could not predict direct CP asymmetries in those decay modes. First, the QCD
factorization approach  had been  applied for
the simplest  charmless $B\rightarrow\pi\pi$ and $B\rightarrow\pi K$
decays \cite{Beneke1,Beneke3,Muta,Du1,Du2} then extended to the
vector  and exotic mesons in final states
\cite{Yang,ChengH1,ChengH2,Diehl}  and $\eta$ or $\eta'$ with a
pseudoscalar or vector kaon \cite{Beneke4}. In \cite{Du3,Du4,Sun},
decay modes of $B_{s}$ meson are discussed. A comprehensive study of
the exclusive hadronic B-meson decays into the final states containing two
pseudoscalar mesons (PP) or a pseudoscalar and a vector meson (PV )
is discussed in \cite{Beneke5}. The Charmless anti-$B_s \rightarrow
VV$ decays has  also been analyzed   in QCD factorization in
\cite{Li}. The hard-scattering kernels relevant to the
negative-helicity decay amplitude in B decays to two vector mesons
are calculated in \cite{Beneke6} in the same  framework. The
two-body hadronic decays of B mesons into pseudoscalar and axial
vector mesons have been studied within the framework of QCD
factorization in \cite{Kwei1}. A detailed study of charmless
two-body B decays into final states involving two vector mesons (V V
) or two axial-vector mesons (AA) or one vector and one axial-vector
meson (V A) has also been done within the framework of QCD
factorization in \cite{Kwei2}. Considering the contributions of both
current-current and penguin operators, the amplitudes and  branching
ratios are recently  estimated at the leading approximation for
$B_c\rightarrow B^{* }P,BV$ in \cite{Junfeng}.

In the present work, taking into account the contributions of the  current-current, QCD penguin and electroweak penguin operators at the leading approximation, we describe the charmed decays  $B_q\to D_q(D_q^*) P$ and
$B_q\to D_q(D_q^*) V$ in the framework of the QCD factorization method. First, using the  factorization method, we
calculate the decay amplitudes and decay widths of these decays in
terms of the transition form factors of the $B_q\to D_q$ and  $B_q\to D^{*}_q$. Having calculated these
transition form factors in the framework of the QCD sum rules in our previous works in
\cite{kazem1,kazem2}, we calculate the branching ratio of these
decays. In order to estimate the approximate branching ratios  and to have a sense of the order of
amplitudes, we make a rough approximation, i.e. at the leading order of $\alpha_s$. Within
this approximation, the hard-scattering kernel functions become very simple and equal to unity \cite{Junfeng}. In this approximation, the long-distance interactions between the $P(V)$ and $B_{q} -D_q(D_q^*)$ system could be neglected. The higher order $\alpha_s$ corrections might not be  small, but calculation of these contributions is not as easy as  the light systems in  final state and needs much more efforts. Hence, for obtaining the exact results on the considered transitions, those effects should be encountered in the future works.  There are several methods in which such type contributions can be  studied, QCD Factorization \cite{Beneke1,Beneke2,Beneke3,Beneke4,Beneke5,Beneke6}, Perturbative QCD \cite{pqcd} and  Soft-Collinear Effective Theory \cite{scet1,scet2}. For more detail analysis of NNLO corrections to $B \rightarrow light-light$ systems and  higher order QCD corrections to the charmless B decays see also \cite {bell1,bell2,bell3,bell4,bell5,philip}. Note that, some of the considered  decays in this paper  have been analyzed  in the
framework of the perturbative QCD  in (PQCD) \cite{pqcd} and for some of them, we have some experimental data  \cite{Yao}.

The outline
of the paper is as follows: In section 2, we calculate the decay
amplitudes and decay widths for $B_q\to D_q(D_q^*) P$ and $B_q\to
D_q(D_q^*) V$ transitions.  Finally, section 3
is devoted to the numerical analysis, a comparison of our results
with the predictions of the PQCD as well as the existing
experimental data and discussion.

\section{Decay amplitudes and decay widths}

In the present  section, we study the  decay amplitudes and decay
widths for $B_q\to D_q(D_q^*) P$ and $B_q\to D_q(D_q^*) V$ decays,
where $q=u, d ,s$ ,~ $P=\pi ,K, D_{q^{'}}$ ~and $V= K^*
,D^{*}_{q^{'}}$~($ q^{'}=d ,s$). At the quark level, the effective
Hamiltonian for $B_q\to D_q(D_q^*)\pi(K,K^*) $  is given by
\begin{eqnarray}\label{hamiltonian}
H_{eff} &=&{G_F\over \sqrt{2}} \left \{ V_{cb} V^*_{uq^{'}}(C_1
O_1^u +C_2 O_2^u ) \right \}.
\end{eqnarray}
Here $O_1^{u}$~and $O_2^{u}$  are quark  operators and are given by
\begin{equation}\begin{array}{llllll}
O_{1}^u&=&(\bar{q_{i}^{\prime
}}u_{i})_{V-A}(\bar{c}_{j}b_{j})_{V-A},  &
O_{2}^u&=&(\bar{q_{i}^{\prime }}u_{j})_{V-A}(\bar{c}
_{j}b_{i})_{V-A},
\end{array}\end{equation}
where $q^{'}=d,s$ and $(\bar q_{1}q_{2})_{V\pm A}=\bar
q_{1}\gamma^{\mu}(1\pm\gamma_{5})q_{2}$. However, the effective Hamiltonian
for $B_q\to D_q(D_q^*)~D_{q^{'}}$ ~and $B_q\to
D_q(D_q^*)~D_{q^{'}}^{*} $~ at the quark level can be written as
\begin{eqnarray}\label{hamiltonian}
H_{eff} &=&{G_F\over \sqrt{2}} \left \{ V_{cb} V^*_{cq^{'}}(C_1
O_1^c +C_2 O_2^c ) +  V_{tb} V^*_{tq^{'}} \sum_{n=3}^{10} C_n
O_n\right \}.
\end{eqnarray}
Here $O_n$ are quark  operators and are given by
\begin{equation}\begin{array}{llllll}
O_{1}^c&=&(\bar{q_{i}^{\prime
}}c_{i})_{V-A}(\bar{c}_{j}b_{j})_{V-A},  &
O_{2}^c&=&(\bar{q_{i}^{\prime }}c_{j})_{V-A}(\bar{c}
_{j}b_{i})_{V-A},   \\\\
O_{3(5)}&=&(\bar{q_{i}^{\prime
}}b_{i})_{V-A}\sum_{q}(\bar{q}_{j}q_{j})_{V-(+)A},  &
O_{4(6)}&=&(\bar{q_{i}^{\prime }}b_{j})_{V-A}\sum_{q}(\bar{q}
_{j}q_{i})_{V-(+)A},   \\\\
O_{7(9)}&=&(\bar{q_{i}^{\prime}}b_{i})_{V-A}\sum_{q}{\frac{3}{2}}e_{q}(\bar{q}
_{j}q_{j})_{V+(-)A},  & O_{8(10)}&=&(\bar{q_{i}^{\prime
}}b_{j})_{V-A}\sum_{q}{\frac{3}{2}}e_{q}(\bar{q}_{j}q_{i})_{V+(-)A},
\end{array}\end{equation}

 where $\sum_q $ sums over $q=u,d,c,s,b$ and
$i$ and $j$ are color indices. The operators $O_{1}$ and $O_{2}$
are called the current-current operators, $O_{3}$...$O_{6}$ are
QCD penguin operators and $O_{7}$...$O_{10}$ are called the
electroweak penguin operators.

The Wilson coefficients $C_n$ have been calculated in different
schemes \cite{BB1,BB2,BB3,BB4}. In this paper we will use consistently the naive
dimensional regularization (NDR) scheme. The values of $C_n$ at $\mu
\approx m_b$ with the next-to-leading order (NLO) QCD corrections
are given by \cite{BB1,BB2,BB3,BB4}

\begin{equation}\begin{array}{llllll}
 C_1 & = & 1.117\ ,  &  C_2 & = & - 0.257\ , \\
 C_3 & = & 0.017\ ,  &  C_4 & = & -0.044\ , \\
 C_5 & = & 0.011\ ,  &  C_6 & = & -0.056\ , \\
 C_7 & = & -1\times 10^{-5}\ , &  C_8 & = & 5\times 10^{-4}\ , \\
 C_9 & = & -0.010\ ,  & C_{10} & = & 0.002\ . \\
\end{array}\end{equation}

The decay width and the branching ratio of the nonleptonic process
$B_q\rightarrow D_q({D^{*}}_{q})M$, where $M$ stands for the $P$
or $V$ mesons, is given by:
\begin{eqnarray}
\Gamma(B_q\rightarrow D_q({D^{*}}_{q})M)&=&{1\over16\pi
m_{B_q}^3}|{\cal A}|^2 \sqrt{\lambda(m_{B_q}^
2,m^2_{D_{q}(D^{*}_{q})},m_M^2)}\,\nonumber\\
Br(B_q\rightarrow D_q({D^{*}}_{q})M)&=&
\tau_{B_q}{\Gamma(B_q\rightarrow D_q({D^{*}}_{q})M)}
\end{eqnarray}
where, $\lambda(x,y,z)=x^2+y^2+z^2-2xy-2xz-2yz$ is usual triangle function.
\\\\
To obtain the decay width, we should calculate the amplitude
$\mathcal{A}$. This amplitude is obtained using the
 factorization method and the definition of the related matrix elements in terms of
 form factors for the $B_q\to D_q$ and $B_q\to D_q^*$ weak transitions as:
\begin{equation}  \label{e3}
<D_q(p^{\prime}){\mid }\bar{c}{\gamma }_{\mu }(1-{\gamma }_{5})b{\mid }%
B_{q}(p)>=(p+p')_{\mu }f^{B_q\to D_q}_{+}(q^{2})+~{(p-p')}_{\mu
}f^{B_q\to D_q}_{-}(q^{2}),
\end{equation}

\begin{equation}\label{3au}
<D_q^*(p',\varepsilon)\mid\overline{c}\gamma_{\mu} b\mid
B_q(p)>=\frac{-2f^{B_q\to
D^*_q}_{1}(q^2)}{m_{D_q^*}+m_{B_{q}}}\varepsilon_{\mu\nu\alpha\beta}
\varepsilon^{\ast\nu}p^\alpha p'^\beta
\end{equation}
\begin{eqnarray}\label{4au}
< D_q^*(p',\varepsilon)\mid\overline{c}\gamma_{\mu}\gamma_{5}
b\mid B_{q}(p)> &=&-i\left[f^{B_q\to
D^*_q}_{0}(q^2)(m_{D_q^*}+m_{B_{q}})\varepsilon_{\mu}^{\ast}
\right. \nonumber \\
+ \frac{f^{B_q\to
D^*_q}_{2}(q^2)}{m_{D_q^*}+m_{B_{q}}}(\varepsilon^{\ast}p)(p+p')_{\mu}
&+& \left. \frac{f^{B_q\to
D^*_q}_{3}(q^2)}{m_{D_q^*}+m_{B_{q}}}(\varepsilon^{\ast}p)(p-p')_{\mu}\right],
\end{eqnarray}
where $q^2$ is transferred momentum square, $q^2=(p-p')^2$, and $p$ and $p'$ are momentum of the initial and final meson states, respectively.

We obtain the ~$\mathcal{A}$ as following:

for $B_q\to D_q P$ ($ P=\pi , K$)~ and~ $B_q\to D_q D_{q'} $:
\begin{eqnarray}\label{re1}
\mathcal{A}^{B_{d(s)}\rightarrow D_{d(s)}P} &=&i\frac{G_{F}}{\sqrt{2}}%
~V_{cb}V_{uq'}^{\ast
}~a_{1}~f_{P}~F_1^{B_{d(s)}\rightarrow D_{d(s)}}(m_{P}^{2}), \nonumber\\
\mathcal{A}^{B_{u}\rightarrow D_{u}P} &=&i\frac{G_{F}}{\sqrt{2}}%
~V_{cb}V_{uq^{^{\prime }}}^{\ast}\Big[
a_{1}~f_{P}~F_1^{B_{u}\rightarrow D_{u}}(m_{P}^{2})
+a_{2}~f_{D_u}~F_1^{B_{u}\rightarrow P}(m_{D_u}^{2})\Big]\nonumber\\
\mathcal{A}^{B_{q}\rightarrow D_{q}D_{q'}} &=&i\frac{G_{F}}{%
\sqrt{2}}~f_{{D_{q'}}%
}~F_1^{B_{q}\rightarrow
D_{q}}(m_{D_{q'}}^{2})~[V_{cb}V_{cq'}^{\ast
}~a_{1}~-V_{{tb}}V_{tq'}^{\ast }(a_{4}+a_{10} \nonumber\\
&&+R_{q'}(a_{6}+a_{8}))],\nonumber\\
\end{eqnarray}
where,
\begin{eqnarray}\label{re11}
F_1^{B_{q}\rightarrow P_1}(m_{P_2}^{2})
&=&(m_{B_{q}}^{2}-m_{P_1}^{2})f_{+}^{B_{q}\rightarrow
P_1}(m_{P_2}^{2})+m_{P_2}^{2}f_{-}^{B_{q}\rightarrow
P_1}(m_{P_2}^{2}),\nonumber\\
R_{q^{\prime }} &=&\frac{2m_{D_{q^{^{\prime }}}}^{2}}{(m_{b}-m_{c})(m_{q^{^{%
\prime }}}+m_{c})},
\end{eqnarray}

for $B_q\to D_q K^{*}$~and~$B_q\to D_q D^{*}_{q'}$ :
\begin{eqnarray}\label{re2}
\mathcal{A}^{B_{d(s)}\rightarrow D_{d(s)}K^{\ast }} &=&
\frac{2~m_{{K}^{\ast }}~f_{{K}^{\ast }}G_{F}}{\sqrt{2}}%
~V_{cb}V_{us}^{\ast }~a_{1}~( ~p~.\varepsilon_{K^{\ast}}^{\ast}
)f_{+}^{B_{d(s)}\rightarrow D_{d(s)}}(m_{K^{\ast }}^{2}), \nonumber \\
\mathcal{A}^{B_{u}\rightarrow D_{u}K^{\ast }} &=&\frac{G_{F}}{\sqrt{2}}%
~V_{cb}V_{us}^{\ast }~(p~.\varepsilon_{K^{\ast}}^\ast )\Big[ 2
a_{1}~m_{{K}^{\ast }}~f_{{K}^{\ast }}~f_{+}^{B_{u}\rightarrow
D_{u}}(m_{K^{\ast }}^{2})- a_2~f_{D_u}
F_2^{B_{u}\rightarrow K^{\ast }}(m_{D_u}^2) \Big], \nonumber \\
\mathcal{A}^{B_{q}\rightarrow D_{q}D_{q'}^{\ast }}
&=&\frac{2~m_{{D_{q'}^{\ast }}}~f_{{D_{q'}^{\ast }}}G_{F}%
}{\sqrt{2}}%
~(p~.\varepsilon_{D_{q'}^{\ast}} ^{\ast })f_{+}^{B_{q}\rightarrow
D_{q}}(m_{D_{q'}^{\ast }}^{2})[V_{cb}V_{cq^{^{\prime }}}^{\ast
}~a_{1} -V_{{tb}}V_{tq^{^{\prime }}}^{\ast
}(a_{4}+a_{10})],\nonumber\\
\end{eqnarray}
where,
\begin{eqnarray}
F_2^{B_{q}\rightarrow V}(m_{P}^2)&=& f_0^{B_{q}\rightarrow
V}(m_{P}^2)(m_{B_q}+m_V)+f_2^{B_{q}\rightarrow
V}(m_{P}^2)(m_{B_q}-m_V) +\frac{f_3^{B_{q}\rightarrow
V}(m_{P}^2)}{(m_{B_q}+m_V)}m_{P}^2,\nonumber\\
\end{eqnarray}

for $B_q\to D^{*}_q P$($ P=\pi , K$)~,and~$B_q\to D^{*}_q
D_{q^{'}}$:
\begin{eqnarray}\label{re3}
\mathcal{A}^{B_{q}\rightarrow D_{d(s)}^{\ast }P} &=&-\frac{G_{F}}{\sqrt{2}}%
~V_{cb}V_{uq^{^{\prime }}}^{\ast
}~a_{1}~f_{P}~(p~.\varepsilon_{D^\ast} ^{\ast
})~F_2^{B_{q}\rightarrow D_{d(s)}^{\ast }}(m_{P}^{2}),\nonumber \\
\mathcal{A}^{B_{u}\rightarrow D_{u}^{\ast }P} &=&-\frac{G_{F}}{\sqrt{2}}%
~V_{cb}V_{uq^{^{\prime }}}^{\ast}(p~.\varepsilon_{D^\ast} ^{\ast
}) ~[a_{1}~f_{P}~F_2^{B_{u}\rightarrow D_{u}^{\ast }}(m_{P}^{2})
- 2a_2~ f_{D_u^\ast}~f_{+}^{B_u \rightarrow P}(m_{D_{u}^\ast}^2)],\nonumber \\
\mathcal{A}^{B_{q}\rightarrow D_{q}^{\ast }D_{q^{^{\prime }}}} &=&-\frac{G_{F}%
}{\sqrt{2}}~f_{{D_{q^{^{\prime }}}}}~(p~.\varepsilon_{D_q^\ast}
^{\ast })~F_2^{B_{q}\rightarrow D_{q}^{\ast }}(m_{D_{q^{\prime
}}}^{2})[V_{cb}V_{cq^{^{\prime }}}^{\ast
}~a_{1}~-V_{{tb}}V_{tq^{^{\prime
}}}^{\ast }(a_{4}+a_{10}\nonumber \\
&&+R_{q^{^{\prime }}}^{^{\prime }}(a_{6}+a_{8}))],
\end{eqnarray}
with
\begin{eqnarray}\label{re111}
R_{q^{^{\prime }}}^{^{\prime }} &=&-\frac{2m_{D_{q'}}^{2}}{%
(m_{b}+m_{c})(m_{q'}+m_{c})},
\end{eqnarray}

and for $B_q\to D^{*}_q K^{*}$~and~$B_q\to D^{*}_q D^{*}_{q^{'}}$
:

\begin{eqnarray}\label{re4}
\mathcal{A}^{B_{d(s)}\rightarrow D_{d(s)}^{\ast }K^{\ast }}
&=&i\frac{m_{{K^{\ast }}}~f_{{K^{\ast }}}G_{F}}{%
\sqrt{2}}~V_{cb}V_{us}^{\ast }~a_{1}~%
~\left[F_3^{B_{d(s)}\rightarrow D_{d(s)}^{\ast }}(m_{{K^{\ast
}}}^2)~(\varepsilon^{\ast}_{D^{\ast}}.\varepsilon^{\ast}_{K^{\ast}}
)\right. \nonumber\\
&&\left.+  F_4^{B_{d(s)}\rightarrow D_{d(s)}^{\ast }}(m_{{K^{\ast
}}}^2)~(p~.\varepsilon^{\ast} _{D^{\ast}})(%
p~.\varepsilon^{\ast}_{K^{\ast}} )-i F_5^{B_{d(s)}\rightarrow
D_{d(s)}^{\ast }}(m_{{K^{\ast }}}^2) \varepsilon _{\mu \nu \rho
\sigma }\varepsilon^{\ast \mu} _{K^{\ast}}
\varepsilon_{D^\ast} ^{\ast \nu }p^{\rho }p_{D^\ast}%
^\sigma \right],\nonumber \\
\mathcal{A}^{B_{u}\rightarrow D_{u}^{\ast }K^{\ast }} &=&i \frac{G_{F}}{%
\sqrt{2}}~V_{cb}V_{us}^{\ast }\Bigg(~a_{1}~m_{{K^{\ast }}}~f_{{K^{\ast }}%
}~\left[F_3^{B_{u}\rightarrow D_{u}^{\ast }}(m_{{K^{\ast
}}}^2)~(\varepsilon^{\ast}_{D^{\ast}}.\varepsilon^{\ast}_{K^{\ast}}
)\right. \nonumber\\
&&\left.+ F_4^{B_{u}\rightarrow D_{u}^{\ast }}(m_{{K^{\ast
}}}^2)~(p~.\varepsilon^{\ast} _{D^{\ast}})(%
p~.\varepsilon^{\ast}_{K^{\ast}} )-i F_5^{B_{u}\rightarrow
D_{u}^{\ast }}(m_{{K^{\ast }}}^2) \varepsilon _{\mu \nu \rho
\sigma }\varepsilon^{\ast \mu} _{K^{\ast}} \varepsilon_{D^\ast}
^{\ast \nu }p^{\rho }p_{D^\ast}%
^\sigma\right] \nonumber \\
&& +a_{2}~m_{{D^{\ast }}}~f_{{D^{\ast }}%
}~\left[F_3^{B_{u}\rightarrow K^{\ast }}(m_{{D^{\ast
}}}^2)~(\varepsilon^{\ast}_{K^{\ast}}.\varepsilon^{\ast}_{D^{\ast}}
) + F_4^{B_{u}\rightarrow K^{\ast }}(m_{{D^{\ast
}}}^2)~(p~.\varepsilon^{\ast} _{K^{\ast}})(%
p~.\varepsilon^{\ast}_{D^{\ast}} )\right.\nonumber\\
&&\left.-i F_5^{B_{u}\rightarrow K^{\ast }}(m_{{D^{\ast }}}^2)
\varepsilon _{\mu \nu \rho \sigma }\varepsilon^{\ast \mu}
_{D^{\ast}} \varepsilon_{K^\ast} ^{\ast
\nu }p^{\rho }{p_{K^\ast}}%
^{\sigma }\right]\Bigg),\nonumber\\
\mathcal{A}^{B_{q}\rightarrow D_{q}^{\ast }D_{q'}^{\ast }} &=&%
i\frac{G_{F}}{\sqrt{2}}~m_{{D_{q'}^{\ast }}}~f_{{D_{q'}^{\ast
}}}~\left[F_3^{B_{q}\rightarrow D_{q}^{\ast }}(m_{{K^{\ast
}}}^2)~(\varepsilon^{\ast}_{D_q^{\ast}}.\varepsilon^{\ast}_{D_{q'}^{\ast}}
)\right. \nonumber\\
&&\left.+ F_4^{B_{q}\rightarrow D_{q}^{\ast }}(m_{{D_{q'}^{\ast
}}}^2)~(p~.\varepsilon^{\ast} _{D_q^{\ast}})(%
p~.\varepsilon^{\ast}_{D_{q'}^{\ast}})-i F_5^{B_{q}\rightarrow
D_{q}^{\ast }}(m_{{D_{q'}^{\ast }}}^2) \varepsilon _{\mu \nu \rho
\sigma }\varepsilon^{\ast \mu} _{D_{q'}^{\ast}}
\varepsilon_{D_q^\ast} ^{\ast \nu }p^{\rho }p_{D_q^\ast}%
^{\sigma }\right]\nonumber\\ && \times[V_{cb}V_{cq^{^{\prime
}}}^{\ast }~a_{1}~-V_{{tb}}V_{tq^{^{\prime }}}^{\ast
}(a_{4}+a_{10})],
\end{eqnarray}
where
\begin{eqnarray}
F_3^{B_q\to V_1}(m_{V_2}^2)&=&(m_{B_{q}}+m_{V_1} )f_{0}^{B_q\to
V_1}(m_{V_2}^2),  \nonumber\\
F_4^{B_q\to V_1}(m_{V_2}^2)&=&\frac{2f_{2}^{B_q\to
V_1}(m_{V_2}^2)}{(m_{B_{q}}+m_{V_1})},\nonumber\\
F_5^{B_q\to V_1}(m_{V_2}^2)&=&\frac{-2 f_{1}^{B_q\to
V_1}(m_{V_2}^2)}{(m_{B_{q}}+m_{V_1})}.
\end{eqnarray}
In the above  expressions, the $\varepsilon^{*}$,
${\varepsilon^{'}}^{*}$, $\varepsilon_{K}$ stand for the polarization of the
$D_{q}^{*}$, $D_{q^{'}}^{*}$ and   $ K^*$ mesons, respectively.
The quantities $a_i$, are given in terms of the coefficient~$C_i$,
\begin{equation}
a_i=C_i +\frac{1}{N_c} C_{i+1} ~(i=odd);~~~~ a_i=C_i +\frac{1}{N_c}
C_{i-1} ~(i=even),
\end{equation}
where $i$ runs from $i=1,...,10$ and $N_c$ is number of color in
QCD. In the above equation, the $a_1$ and $a_2 $ are both related to the  coefficients $C_{1,2}$, which are very large comparing with the other Wilson coefficients, but we will keep all coefficients to get ride of further approximation.

Now we can calculate the decay widths for  $B_q\to D_q(D_q^*) P$ and
$B_q\to D_q(D_q^*) V$ decays. The explicit expressions for decay
widths are given as follow:

\begin{eqnarray}
\Gamma (B_{d(s)} \rightarrow D_{d(s)}P(P=\pi,K))&=&\frac{G_{F}^{2}}{32~\pi m_{B_{d(s)}}^{3}}%
|V_{cb}{V_{uq^{\prime }}^{\ast }}|^{2}~a_1%
^{2}~f_{P}^{2}~\sqrt{\lambda (m_{B_{d(s)}}^{2},m_{D_{d(s)}}^{2},m_{P}^{2})%
}\nonumber\\ &&\times [F_1^{B_{d(s)}\rightarrow
D_{d(s)}}(m_P^2)]^{2},
\end{eqnarray}

\begin{eqnarray}
\Gamma (B_{u} \rightarrow D_{u}P(P=\pi,K))&=&\frac{G_{F}^{2}}{32~\pi m_{B_{u}}^{3}}%
|V_{cb}{V_{uq^{\prime }}^{\ast }}|^{2}\sqrt{\lambda (m_{B_{u}}^{2},m_{D_{u}}^{2},m_{P}^{2})%
} \left[a_1%
^{2}~f_{P}^{2}~[F_1^{B_{u}\rightarrow D_{u}}(m_P^2)]^{2}  \right. \nonumber\\
&&\left.+ a_2%
^{2}~f_{D_u}^{2}~[F_1^{B_{u}\rightarrow
P}(m_{D_u}^2)]^{2}\right.\nonumber\\&&\left. +
2~a_1~a_2~f_{D_u}~f_P~[F_1^{B_{u}\rightarrow
D_{u}}(m_P^2)~F_1^{B_{u}\rightarrow P}(m_{D_u}^2)]~ \right],
\end{eqnarray}

\begin{eqnarray}
\Gamma (B_{q} \rightarrow D_{q}D_{q^{\prime
}})&=&\frac{G_{F}^{2}}{32~\pi m_{B_{q}}^{3}}f_{D_{q^{\prime
}}}^{2}~\sqrt{\lambda (m_{B_{q}}^{2},m_{D_{q}}^{2},m_{D_{q^{\prime
}}}^{2})}[F_1^{B_{q}\rightarrow
D_{q}}(m_{D_{q^{\prime}}}^{2})]^{2} \nonumber\\
&&~\times \left|V_{cb}V_{cq^{\prime }}^{\ast
}~a_{1}~-V_{{tb}}V_{tq^{\prime }}^{\ast
}[a_{4}+a_{10}+R_{q^{\prime}}(a_{6}+a_{8})]\right|^{2},
\end{eqnarray}

\begin{eqnarray}
\Gamma (B_{d(s)} \rightarrow D_{d(s)}K^{\ast
})&=&\frac{G_{F}^{2}}{32~\pi
m_{B_{d(s)}}^{3}}|V_{cb}{V_{us}^{\ast}}|^{2}~a_{1}^{2}~f_{K^{\ast
}}^{2}\lambda (m_{B_{d(s)}}^{2},m_{D_{d(s)}}^{2},m_{K^{\ast
}}^{2})^\frac{3}{2} \nonumber \\
&&\times [f^{B_{d(s)}\to D_{d(s)}}_{+}(m_{K^{\ast }}^{2})]^2,
\end{eqnarray}

\begin{eqnarray}
\Gamma (B_{u} \rightarrow D_{u}K^{\ast
})&=&\frac{G_{F}^{2}}{32~\pi
m_{B_{u}}^{3}}|V_{cb}{V_{us}^{\ast}}|^{2} \frac{\lambda
(m_{B_{u}}^{2},m_{D_{u}}^{2},m_{K^{\ast
}}^{2})^\frac{3}{2}}{4m_{K^{\ast
}}^{2}}\Big[4~a_{1}^{2}~m_{K^{\ast
}}^{2}~f_{K^{\ast }}^{2}~[f^{B_{u}\to D_{u}}_{+}(m_{K^{\ast}}^{2})]^2\nonumber\\
&& + a_{2}^{2}~f_{D_u}^{2}~{[F_2^{B_{u}\rightarrow
K^{\ast}}(m_{D_u}^{2})]}^2\nonumber\\
&& - 4~a_1~a_2~m_{K^{\ast}}~f_{K^{\ast }}f_{D_u}~f^{B_u\to
D_u}_{+}(m_{K^{\ast}}^{2})~F_2^{B_{u}\rightarrow
K^{\ast}}(m_{D_u}^{2}) \Big],
\end{eqnarray}

\begin{eqnarray}
\Gamma (B_{q} \rightarrow D_{q}D_{q'}^{\ast })&=&\frac{G_{F}^{2}%
}{32~\pi m_{B_{q}}^{3}}f_{D_{q'}^{\ast }}^{2}\lambda
(m_{B_{q}}^{2},m_{D_{q}}^{2},m_{D_{q'}^{\ast
}}^{2})^\frac{3}{2}~[f^{B_q\to D_q}_{+}(m_{D_{q'}^{\ast
}}^{2})]^2\nonumber \\
&&\times \left|V_{cb}V_{cq^{^{\prime }}}^{\ast
}~a_{1}~-V_{{tb}}V_{tq^{^{\prime }}}^{\ast
}[a_{4}+a_{10}]\right|^{2},
\end{eqnarray}

\begin{eqnarray}
\Gamma (B_{d(s)} \rightarrow D_{d(s)}^{\ast }K^{\ast
})&=&\frac{G_{F}^{2}}{32~\pi
m_{B_{d(s)}}^{3}}|{V_{cb}}{V_{us}^{\ast}}|^{2}~ a_{1}^2~m_{K^{\ast
}}^2f_{K^{\ast }}^2~\lambda (m_{B_{d(s)}}^{2},m_{D_{d(s)}^{\ast
}}^{2},m_{K^{\ast
}}^{2})^{\frac{1}{2}}~\nonumber \\
&&\Bigg(\Big[ F_3^{B_{d(s)}\rightarrow D_{d(s)}^{\ast}}(m_{K^{\ast
}}^{2})
\Big]^2 \Big[\frac{%
\lambda (m_{B_{{d(s)}}}^{2},m_{D_{{d(s)}}^{\ast }}^{2},m_{K^{\ast }}^{2})}{%
4m_{D_{{d(s)}}^{\ast }}^{2}m_{K^{\ast }}^{2}}+3 \Big]\nonumber \\
&&+\Big[ F_4^{B_{d(s)}\rightarrow D_{{d(s)}}^{\ast}}(m_{K^{\ast
}}^{2})\Big]^2
\Big[\frac{%
\lambda (m_{B_{{d(s)}}}^{2},m_{D_{{d(s)}}^{\ast }}^{2},m_{K^{\ast }}^{2})^2}{%
16m_{D_{{d(s)}}^{\ast }}^{2}m_{K^{\ast }}^{2}} \Big] \nonumber\\
&&+\Big[ F_5^{B_{d(s)}\rightarrow D_{{d(s)}}^{\ast}}(m_{K^{\ast
}}^{2}) \Big]^2
\Big[\frac{%
\lambda (m_{B_{{d(s)}}}^{2},m_{D_{{d(s)}}^{\ast }}^{2},m_{K^{\ast }}^{2})}{%
2}\Big]\nonumber \\
&&+\Big[ F_3^{B_{d(s)}\rightarrow D_{{d(s)}}^{\ast}}(m_{K^{\ast
}}^{2})\Big] \Big[ F_4^{B_{d(s)}\rightarrow D_{{d(s)}}^{\ast}}(m_{K^{\ast }}^{2})\Big]\nonumber \\
&& \times(m_{B_{d(s)}}^2-m_{D_{d(s)}^{\ast }}^2-m_{K^{\ast }}^2)\Big[ \frac{%
\lambda (m_{B_{{d(s)}}}^{2},m_{D_{{d(s)}}^{\ast }}^{2},m_{K^{\ast }}^{2})}{%
4m_{D_{{d(s)}}^{\ast }}^{2} m_{K^{\ast }}^{2}}\Big]\Bigg),
\end{eqnarray}

\begin{eqnarray}
\Gamma (B_{u} \rightarrow D_{u}^{\ast }K^{\ast
})&=&\frac{G_{F}^{2}}{32~\pi
m_{B_{u}}^{3}}|{V_{cb}}{V_{us}^{\ast}}|^{2}~\lambda
(m_{B_{u}}^{2},m_{D_{u}^{\ast }}^{2},m_{K^{\ast
}}^{2})^{\frac{1}{2}}~ \Bigg(\Big[ a_{1}~m_{K^{\ast }}f_{K^{\ast
}}F_3^{B_u\rightarrow D_{u}^{\ast}}(m_{K^{\ast }}^{2})\nonumber \\
&&+ a_{2}~m_{D_u^{\ast }}f_{D_u^{\ast }}F_3^{B_u\rightarrow
K^{\ast}}(m_{D_u^{\ast
}}^{2})\Big]^2 \Big[\frac{%
\lambda (m_{B_{u}}^{2},m_{D_{u}^{\ast }}^{2},m_{K^{\ast }}^{2})}{%
4m_{D_{u}^{\ast }}^{2}m_{K^{\ast }}^{2}}+3 \Big]\nonumber \\
&&+\Big[ a_{1}~m_{K^{\ast }}f_{K^{\ast }}F_4^{B_u\rightarrow
D_{u}^{\ast}}(m_{K^{\ast }}^{2})+  a_{2}~m_{D_u^{\ast
}}f_{D_u^{\ast }}F_4^{B_u\rightarrow K^{\ast}}(m_{D_u^{\ast
}}^{2})\Big]^2 \nonumber\\
&& \times \Big[\frac{%
\lambda (m_{B_{u}}^{2},m_{D_{u}^{\ast }}^{2},m_{K^{\ast }}^{2})^2}{%
16m_{D_{u}^{\ast }}^{2}m_{K^{\ast }}^{2}} \Big]+\Big[
a_{1}~m_{K^{\ast }}f_{K^{\ast }}F_5^{B_u\rightarrow
D_{u}^{\ast}}(m_{K^{\ast }}^{2})  \nonumber\\ && +
a_{2}~m_{D_u^{\ast }}f_{D_u^{\ast }}F_5^{B_u\rightarrow
K^{\ast}}(m_{D_u^{\ast }}^{2})\Big]^2
\Big[\frac{%
\lambda (m_{B_{u}}^{2},m_{D_{u}^{\ast }}^{2},m_{K^{\ast }}^{2})}{%
2}\Big]\nonumber \\
&&+\Big[ a_{1}~m_{K^{\ast }}f_{K^{\ast }}F_3^{B_u\rightarrow
D_{u}^{\ast}}(m_{K^{\ast }}^{2})+  a_{2}~m_{D_u^{\ast
}}f_{D_u^{\ast }}F_3^{B_u\rightarrow K^{\ast}}(m_{D_u^{\ast
}}^{2})\Big] \nonumber\\
&& \times\Big[ a_{1}~m_{K^{\ast }}f_{K^{\ast }}F_4^{B_u\rightarrow
D_{u}^{\ast}}(m_{K^{\ast }}^{2})+ a_{2}~m_{D_u^{\ast
}}f_{D_u^{\ast }}F_4^{B_u\rightarrow K^{\ast}}(m_{D_u^{\ast
}}^{2})\Big]\nonumber \\
&& \times(m_{B_u}^2-m_{D_u^{\ast }}^2-m_{K^{\ast }}^2)\Big[ \frac{%
\lambda (m_{B_{u}}^{2},m_{D_{u}^{\ast }}^{2},m_{K^{\ast }}^{2})}{%
4m_{D_{u}^{\ast }}^{2} m_{K^{\ast }}^{2}}\Big]\Bigg),
\end{eqnarray}

\begin{eqnarray}
\Gamma (B_{q} \rightarrow D_{q}^{\ast }D_{q'}^{\ast
})&=&\frac{G_{F}^{2}}{32~\pi m_{B_{q}}^{3}}~
m_{D^{\ast}_{q'}}^2f_{D^{\ast}_{q'}}^2~\lambda
(m_{B_{q}}^{2},m_{D_{q}^{\ast }}^{2},m_{D^{\ast}_{q'}
}^{2})^{\frac{1}{2}}~\nonumber \\
&&\Bigg(\Big[ F_3^{B_{q}\rightarrow D_{q}^{\ast}}(m_{D^{\ast}_{q'}
}^{2})
\Big]^2 \Big[\frac{%
\lambda (m_{B_{{q}}}^{2},m_{D_{{q}}^{\ast }}^{2},m_{D^{\ast}_{q'}}^{2})}{%
4m_{D_{{q}}^{\ast }}^{2}m_{D^{\ast}_{q'}}^{2}}+3 \Big]\nonumber \\
&&+\Big[ F_4^{B_{q}\rightarrow D_{{q}}^{\ast}}(m_{D^{\ast}_{q'}
}^{2})\Big]^2
\Big[\frac{%
\lambda (m_{B_{{q}}}^{2},m_{D_{{q}}^{\ast }}^{2},m_{D^{\ast}_{q'}}^{2})^2}{%
16m_{D_{{q}}^{\ast }}^{2}m_{D^{\ast}_{q'}}^{2}} \Big] \nonumber\\
&&+\Big[ F_5^{B_{q}\rightarrow
D_{{q}}^{\ast}}(m_{D^{\ast}_{q'}}^{2}) \Big]^2
\Big[\frac{%
\lambda (m_{B_{{q}}}^{2},m_{D_{{q}}^{\ast }}^{2},m_{D^{\ast}_{q'}}^{2})}{%
2}\Big]\nonumber \\
&&+\Big[ F_3^{B_{q}\rightarrow D_{{q}}^{\ast}}(m_{D^{\ast}_{q'}
}^{2})\Big] \Big[ F_4^{B_{q}\rightarrow D_{{q}}^{\ast}}(m_{D^{\ast}_{q'}}^{2})\Big]\nonumber \\
&& \times(m_{B_{q}}^2-m_{D_{q}^{\ast }}^2-m_{D^{\ast}_{q'}}^2)\Big[ \frac{%
\lambda (m_{B_{{q}}}^{2},m_{D_{{q}}^{\ast }}^{2},m_{D^{\ast}_{q'}}^{2})}{%
4m_{D_{{q}}^{\ast }}^{2}
m_{D^{\ast}_{q'}}^{2}}\Big]\Bigg)\nonumber\\
&& \times \left|V_{cb}V_{cq^{^{\prime }}}^{\ast
}~a_{1}~-V_{{tb}}V_{tq^{^{\prime }}}^{\ast
}[a_{4}+a_{10}]\right|^{2},
\end{eqnarray}

\begin{eqnarray}
\Gamma (B_{d(s)} \rightarrow D_{d(s)}^{\ast
}P(P=\pi,K))&=&\frac{G_{F}^{2}}{128~\pi
m_{B_{d(s)}}^{3}m_{D_{d(s)}^{\ast }}^{2}}|V_{cb} {V_{uq^{\prime
}}^{\ast}}|^{2}~a_{1}^{2}~f_{P}^{2}\lambda
(m_{B_{d(s)}}^{2},m_{D_{d(s)}^{\ast
}}^{2},m_{P}^{2})^\frac{3}{2}\nonumber\\
&&[F_2^{B_{d(s)}\rightarrow D_{d(s)}^{\ast }}(m_P^{2})]^{2},
\end{eqnarray}

\begin{eqnarray}
\Gamma (B_{u} &\rightarrow& D_{u}^{\ast
}P(P=\pi,K))=\frac{G_{F}^{2}}{32~\pi m_{B_{u}}^{3}}|V_{cb}
{V_{uq^{\prime }}^{\ast}}|^{2}\frac{\lambda
(m_{B_{u}}^{2},m_{D_{u}^{\ast
}}^{2},m_{P}^{2})^\frac{3}{2}}{4m_{D_{u}^{\ast
}}^{2}}\nonumber\\
&&\times\left(4~a_{2}^{2}~m_{D_{u}^{\ast }}^{2}~f_{D_{u}^{\ast}
}^{2}~[f^{B_u\to P}_{+}(m_{D_{u}^{\ast }}^{2})]^2  +
a_{1}^{2}~f_{P }^{2}~{[F_2^{B_{u}\rightarrow
D_u^{\ast}}(m_{P}^{2})]}^2\right.\nonumber\\
&&\left. - 4~a_1~a_2~m_{D_u^{\ast}}~f_{P }f_{D_u^{\ast}}~f^{B_u\to
P}_{+}(m_{D_u^{\ast}}^{2})~F_2^{B_{u}\rightarrow
D_u^{\ast}}(m_P^{2}) \right) ,
\end{eqnarray}

\begin{eqnarray}
\Gamma (B_{q} \rightarrow D_{q}^{\ast }D_{q'})&=&\frac{G_{F}^{2}%
}{128~\pi m_{B_{q}}^{3}m_{D_{q}^{\ast }}^{2}}f_{D_{q^{^{\prime
}}}}^{2}\lambda (m_{B_{q}}^{2},m_{D_{q}^{\ast
}}^{2},m_{D_{q^{^{\prime
}}}}^{2})^\frac{3}{2}~[F_2^{B_{q}\rightarrow D_{q}^{\ast }}(m_{D_{q'}}^{2})]^{2} \nonumber\\
&&~\times\left|V_{cb}V_{cq^{^{\prime }}}^{\ast
}~a_{1}~-V_{{tb}}V_{tq'}^{\ast
}[a_{4}+a_{10}+R_{q'}^{\prime}(a_{6}+a_{8})]\right|^{2}.
\end{eqnarray}

\begin{table}
\begin{center}
\label{table5}
\begin{tabular} {|c|c|c|c|c|c|}
\hline $m_\pi$ & $m_ K$ & $m_{D^\pm}$ & $m _{\bar{D}^{0}}$  &
$m_{D_s }$&$m_{K^*(892)}$    \\
\hline $139.570$  & $493.677\pm0.016$    &
$1869.62\pm0.20$ & $1864.84\pm0.17$   &$1968.49\pm0.34$& $891.66\pm0.26$   \\
\hline $m_{{D^*}^\pm}$   & $m _{\bar{D^*}^{0}}$ & $m_{{D^*}_s }$ &
$m_{B^{\pm} }$& $m_{B^{0} }$& $m_{B_s }$   \\
\hline $2010.27\pm0.17$ &  $2006.97\pm0.19$   & $2112.3\pm0.5$  &
$5279.2\pm0.3$ &  $5279.5\pm0.3$ &$5366.3\pm0.6$  \\
\hline
\end{tabular}\end{center}
\caption{Values of the  masses in MeV \cite {Yao}. }
\end{table}

\begin{table}
\begin{center}
\label{table5}
\begin{tabular} {|c|c|c|c|c|c|c|}
\hline $f_\pi\cite{Yao}$ & $f_ K\cite{Yao}$ &
$f_{D^\pm}\cite{Yao}$ & $f _{\bar{D}^{0}}\cite{Yao}$ &
$f_{D_s }\cite{Yao}$ & $f_{K^*}\cite{NS97}$ \\
\hline $130.7\pm0.46$  & $159.8\pm1.84$    &  $222.6\pm19.5$ &
$222.6\pm19.5$ & $294\pm27$
&    $217\pm5$  \\
\hline $f_{{D^*}^\pm}\cite{Yao}$   & $f
_{\bar{D^*}^{0}}\cite{Yao}$ & $f_{{D^*}_s }\cite{Colang}$ &
$f_{B^{\pm} }\cite{Yao}$& $f_{B^{0} }\cite{Yao}$& $f_{B_s }\cite{rolf}$   \\
\hline  $230\pm20$&$230\pm20$   & $266\pm32$  & $176\pm42$ &  $176\pm42$ & $206\pm10$  \\
\hline
\end{tabular}\end{center}
\caption{Values of the decay constants in MeV. }
\end{table}

\begin{table}
\begin{center}
\label{table5}
\begin{tabular} {|c|c|c|c|c|c|}
\hline $|V_{ud}|$ & $|V_{us}|$ & $|V_{cd}|$ & $|V_{cs}|$  &
$|V_{cb}|$ \\
\hline
   $0.97377\pm0.00027$  & $0.2257\pm0.0021$    &  $0.230\pm0.011$ & $0.957\pm0.110$   &    $0.0416\pm0.0006$    \\ \hline$|V_{ub}|$ &
 $|V_{td}|$   & $|V_{tb}|$ & $|V_{ts}|$  &  \\ \hline $0.00431\pm0.00030$    &
 $0.0074\pm0.0008$ &  $0.77\pm0.18$   & $0.0406\pm0.0027$ &   \\
\hline
\end{tabular}\end{center}
\caption{Values of the elements of the CKM matrix \cite {Yao}. }
\end{table}

\section{Numerical analysis}
This section encompasses our numerical analysis,  comparison of
our results with the predictions of the PQCD as well as the
existing experimental data and discussion. The expressions of the
amplitudes and decay widths depict that the main input parameters
entering the expressions are  Wilson coefficients presented in the
section 2, elements of the CKM matrix,  leptonic decay constants,
Borel parameters $M_{1}^2$ and $M_{2}^2$ as well as the continuum
thresholds $s_{0}$ and $s'_{0}$ \cite{kazem1,kazem2}. In further
numerical analysis, we choose the  numerical values as presented
in the Tables 1, 2 and 3. The  Borel mass squares $M_{1}^2$ and
$M_{2}^2$ and continuum thresholds $s_{0}$  and $s'_{0}$ are
auxiliary parameters, hence the physical quantities  should be
independent of them. The parameters $s_0$ and $s_0^\prime$, are
 determined from the conditions that
guarantees the sum rules for form factors to have the best
stability in the allowed $M_1^2$ and $M_2^2$ region. The working
regions for $M_1^2$ and $M_2^2$  as well as the values for
continuum thresholds are determined in \cite{kazem1,kazem2}. Here,
we choose the values $s_{0}=35\pm5~GeV^2$,
$s^{'}_{0}=7\pm1~GeV^2$, $M_{1}^2=17.0\pm2.5~GeV^2$,
$M_{2}^2=7\pm1~GeV^2$ from those working values for auxiliary
parameters. The values of the form factors $f_{\pm}$ and
$f_{0,1,2,3}$ at different values of $q^2$ which we need in the
expressions for decay widths  are presented in Tables 4 and 5,
respectively. Using the expressions for total decay widths, the
values of branching fractions for $B_{q}\rightarrow D_{q}P$,
$B_{q}\rightarrow D_{q}^{*}P$, $B_{q}\rightarrow D_{q}V$ and
$B_{q}\rightarrow D_{q}^{*}V$ are found. We depict the  values of
the  branching ratios in Tables 6, 7, 8 and 9.  Here, we should
stress that, as we mentioned before, our results depicted in the
Tables are approximate results since we considered the observable
only at the leading order of $\alpha_s$. To obtain more exact
results the higher order  $\alpha_s$ corrections
should be taken into account. However,  the  presented
uncertainties in the results are belong to the uncertainties in
the values of the input parameters as well as variations in form
factors which are related to the errors in determination of the
 auxiliary parameters namely, Borell mass parameters $M_{1}^2$
and $M_{2}^2$ and continuum thresholds $s_{0}$  and $s'_{0}$.
These Tables also include a comparison of our results with the
existing predictions of the PQCD as well as the experimental data.
From these Tables, we see a good consistency among  two non-perturbative approaches
and the experiment in order of magnitude. In many cases, the presented results out of order of magnitude from three approaches coincide   especially, when we consider the uncertainties
in the results. The best consistency  between our results and predictions of the PQCD is  related to the $B_s^{0}\rightarrow{D_s^*}^{\pm}{K}^{\mp}$ transition and  $B_s^{0}\rightarrow{D_s}^{\pm}{K^{*}(892)}^{\mp}$ transition shows the biggest  discrepancy between two methods. Our central value prediction on $B^{0}\rightarrow\bar{D^*}^{\pm}{K}^{\mp}$ is approximately the same as the experimental result, however, the central experimental result on the branching ratio of $B_s^{0}\rightarrow{D^*_s}^{\pm}{D_s^*}^{\mp}$ depicts a big discrepancy comparing that of our prediction.  The presented predictions from PQCD are related to
the charm-charmless cases in the final states and we have no
predictions on the charm-charm cases from this approach. In this
approach, the wave functions of the participating mesons, which
are available with higher order corrections, have been used to
calculate the amplitudes \cite{pqcd}. Therefor, over all agreement between
our results and predictions of  PQCD for charm-light cases in the final state and   the
experimental data for both charm-light and charm-charm cases, could
be considered as a good test of the QCD factorization at leading
order of $\alpha_s$ for related transitions.  However, for exact comparison, much more
efforts are needed in the future works, which may include the
higher order corrections.  Our  results of some decay modes which
have not been measured in the experiment can be tested in the
future experiments at  LHCb and other B factories.

In conclusion, using the QCD factorization approach and taking into account the contributions of the  current-current, QCD penguin and the electroweak penguin operators at the leading approximation, the decay
amplitudes and decay widths of $B_q\to D_q(D_q^*) P$ and $B_q\to
D_q(D_q^*) V$ transitions were calculated in terms of the transition
form factors of the $B_q\to D_q(D_q^*) $. Having computed those form factors in the framework of the  three-point QCD
sum rules in our previous works, the branching fraction for these decays were also
evaluated. A comparison of our results with the predictions of
the perturbative QCD as well as the existing experimental data was
presented. Our results are over all in a good agreement with the predictions
of   the  PQCD and the existing experimental data. Our predictions on some transitions, which have no experimental data can be checked by future experiments at LHCb or other B factories. To get  more exact results from the QCD factorization method, higher order  $\alpha_s$ corrections  should be considered in the future works.

\section{Acknowledgments}

The authors would like to thank T. M. Aliev and A. Ozpineci for
their useful discussions. One of the authors (K. Azizi) thanks
Turkish Scientific and Research Council (TUBITAK) for their partial
financial support provided under the
project 108T502.

\newpage

\begin{table}[tbp]
\begin{center}
\begin{tabular}{||c|ccccccc||}
\hline  $q^{2}$ &  \lower 0.30cm\hbox{{\vrule width 0pt height
1.cm}}$m_{\pi^{\pm}}^{2} $ & $ m_{K^{\pm}}^{2} $&
$m_{{K^{*}(892)}^{\pm}}^{2}$& $
m_{D^{\pm}}^{2}$& $m_{D_{s}}^{2}$&
$m_{{D^{*}}^{\pm}}^{2}$& $m_{D^{*}_{s}}^{2}$\hbox\\
\hline\hline $~f_{+}^{B^+\rightarrow \bar{D}^0}(q^2)~$ & \lower
0.30cm\hbox{{\vrule width 0pt height
1.cm}} \tiny{$0.59\pm0.14 $ }& \tiny{$0.60\pm0.15$}& \tiny{$0.63\pm0.16$}& \tiny{$0.86\pm0.22$ }& \tiny{$0.92\pm0.23$} & \tiny{$0.96\pm0.24 $} & \tiny{$1.08\pm0.27 $}  \\
\hline $~f_{-}^{B^+\rightarrow \bar{D}^0}(q^2)~$ & \lower
0.30cm\hbox{{\vrule width 0pt height
1.cm}} \tiny{$-0.20\pm0.05$} & \tiny{$-0.21\pm0.05$ }& \tiny{$-0.22\pm0.06$}&\tiny{$-0.38\pm0.10$} &\tiny{$-0.44\pm0.11$} &\tiny{$-0.49\pm0.12$} &\tiny{$-0.69\pm0.17$}  \\
\hline $~f_{+}^{B^0\rightarrow D^+}(q^2)~$ & \lower
0.30cm\hbox{{\vrule width 0pt height
1.cm}} \tiny{$0.58\pm0.15$} &\tiny{ $0.59\pm0.15$} & \tiny{$0.63\pm0.17$}&\tiny{$0.86\pm0.22$ }&\tiny{$0.92\pm0.22$} &\tiny{$0.95\pm0.23$} &\tiny{$1.08\pm0.27$}  \\
\hline $~f_{-}^{B^0\rightarrow D^+}(q^2)~$ &  \lower
0.30cm\hbox{{\vrule width 0pt height
1.cm}} \tiny{$-0.20\pm0.05$ }& \tiny{$-0.21\pm0.05$}&\tiny{$-0.22\pm0.06$} &\tiny{$-0.37\pm0.10$} &\tiny{$-0.44\pm0.11$} &\tiny{$-0.49\pm0.12$} &\tiny{$-0.69\pm0.17$} \\
\hline$~f_{+}^{B_s^0\rightarrow D_s^+}(q^2)~$ &  \lower
0.30cm\hbox{{\vrule width 0pt height
1.cm}} \tiny{$0.26\pm0.06$} & \tiny{$0.27\pm0.06$}&\tiny{$0.28\pm0.07$}&\tiny{$0.35\pm0.09$} &\tiny{$0.38\pm0.10$} &\tiny{$0.39\pm0.10$ }&\tiny{$0.42\pm0.11$} \\
\hline$~f_{-}^{B_s^0\rightarrow D_s^+}(q^2)~$ &  \lower
0.30cm\hbox{{\vrule width 0pt height
1.cm}}\tiny{$-0.11\pm0.03$}&\tiny{$-0.12\pm0.03$}&\tiny{$-0.13\pm0.03$} &\tiny{$-0.15\pm0.04$}
&\tiny{$-0.16\pm0.04$} &\tiny{$-0.17\pm0.05$}&\tiny{$-0.18\pm0.05$}
\\ \hline
\end{tabular}
\vskip 0.3 cm

\end{center}
\caption{The values of form factors $f_{\pm}$  at different values
of $q^2$.}
\end{table}

\begin{table}[tbp]
\begin{center}
\begin{tabular}{||c|ccccccc||}
\hline  $q^{2}$ &  \lower 0.30cm\hbox{{\vrule width 0pt height
1.cm}}$m_{\pi^{\pm}}^{2} $ & $ m_{K^{\pm}}^{2} $&
$m_{{K^{*}(892)}^{\pm}}^{2}$& $
m_{D^{\pm}}^{2}$& $m_{D_{s}}^{2}$&
$m_{{D^{*}}^{\pm}}^{2}$& $m_{D^{*}_{s}}^{2}$\hbox\\
\hline\hline $~f_{0}^{B^+\rightarrow \bar{D^*}^0}(q^2)~$ & \lower
0.30cm\hbox{{\vrule width 0pt height
1.cm}} \tiny{$0.76\pm0.19 $} &\tiny{$0.78\pm0.19$}&\tiny{$0.80\pm0.20$}&\tiny{$0.97\pm0.24$} &\tiny{$1.01\pm0.25$} &\tiny{$1.09\pm0.26 $ }&\tiny{$1.13\pm0.27 $}  \\
\hline $~f_{1}^{B^+\rightarrow \bar{D^*}^0}(q^2)~$ & \lower
0.30cm\hbox{{\vrule width 0pt height
1.cm}} \tiny{$0.62\pm0.15$} & \tiny{$0.63\pm0.15$ }& \tiny{$0.67\pm0.16$}&\tiny{$0.98\pm0.24$ }&\tiny{$1.08\pm0.26$} &\tiny{$1.14\pm0.27$}&\tiny{$1.27\pm0.29$}  \\
\hline $~f_{2}^{B^+\rightarrow \bar{D^*}^0}(q^2)~$ & \lower
0.30cm\hbox{{\vrule width 0pt height
1.cm}}\tiny{ $0.90\pm0.22$} & \tiny{$0.96\pm0.23$ }& \tiny{$0.99\pm0.24$}&\tiny{$1.50\pm0.38$} &\tiny{$1.60\pm0.40$} &\tiny{$1.82\pm0.46$} &\tiny{$2.00\pm0.50$ } \\
\hline $~f_{3}^{B^+\rightarrow \bar{D^*}^0}(q^2)~$ &  \lower
0.30cm\hbox{{\vrule width 0pt height
1.cm}} \tiny{$-1.51\pm0.38$ }& \tiny{$-1.62\pm0.40$}&\tiny{$-1.65\pm0.40$ }& \tiny{$-2.01\pm0.50$} &\tiny{$-2.30\pm0.55$} &\tiny{$-2.50\pm0.60$} &\tiny{$-2.65\pm0.61$} \\
\hline$~f_{0}^{B^0\rightarrow {D^*}^+}(q^2)~$&  \lower
0.30cm\hbox{{\vrule width 0pt height
1.cm}} \tiny{$0.76\pm0.19$} & \tiny{$0.78\pm0.19$}&\tiny{$0.81\pm0.20$}&\tiny{$0.97\pm0.24$}&\tiny{$1.02\pm0.25$} &\tiny{$1.10\pm0.26$} &\tiny{$1.13\pm0.27$} \\
\hline$~f_{1}^{B^0\rightarrow {D^*}^+}(q^2)~$&  \lower
0.30cm\hbox{{\vrule width 0pt height
1.cm}} \tiny{$0.61\pm0.15$} &\tiny{ $0.63\pm0.15$ }& \tiny{$0.66\pm0.16$}&\tiny{$0.98\pm0.24$ }&\tiny{$1.07\pm0.26$} &\tiny{$1.14\pm0.27$ }&\tiny{$1.27\pm0.29$ } \\
\hline$~f_{2}^{B^0\rightarrow {D^*}^+}(q^2)~$ &  \lower
0.30cm\hbox{{\vrule width 0pt height
1.cm}} \tiny{$0.90\pm0.22$ }& \tiny{$0.95\pm0.23$} &\tiny{ $0.99\pm0.24$}&\tiny{$1.51\pm0.38$} &\tiny{$1.61\pm0.40$} &\tiny{$1.82\pm0.46$} &\tiny{$2.01\pm0.50$ } \\
\hline$~f_{3}^{B^0\rightarrow {D^*}^+}(q^2)~$ &  \lower
0.30cm\hbox{{\vrule width 0pt height
1.cm}} \tiny{$-1.52\pm0.38$ }& \tiny{$-1.62\pm0.40$}&\tiny{$-1.66\pm0.40$ }&\tiny{$-2.01\pm0.50$ }&\tiny{$-2.31\pm0.55$} &\tiny{$-2.51\pm0.60$} &\tiny{$-2.65\pm0.61$} \\
\hline$~f_{0}^{B_s^0\rightarrow {D_s^*}^+}(q^2)~$ &  \lower
0.30cm\hbox{{\vrule width 0pt height
1.cm}} \tiny{$0.38\pm0.10$} &\tiny{ $0.40\pm0.11$}&\tiny{$0.41\pm0.11$}&\tiny{$0.58\pm0.15$} &\tiny{$0.62\pm0.16$} &\tiny{$0.67\pm0.17$} &\tiny{$0.70\pm0.18$ }\\
\hline$~f_{1}^{B_s^0\rightarrow {D_s^*}^+}(q^2)~$ &  \lower
0.30cm\hbox{{\vrule width 0pt height
1.cm}} \tiny{$0.33\pm0.08$ }& \tiny{$0.36\pm0.08$}&\tiny{$0.40\pm0.11$}&\tiny{$0.67\pm0.17$ }&\tiny{$0.71\pm0.17$} &\tiny{$0.74\pm0.18$} &\tiny{$0.76\pm0.18$} \\
\hline$~f_{2}^{B_s^0\rightarrow {D_s^*}^+}(q^2)~$ &  \lower
0.30cm\hbox{{\vrule width 0pt height
1.cm}}\tiny{ $0.43\pm0.11$} & \tiny{$0.47\pm0.12$}&\tiny{$0.50\pm0.13$}&\tiny{$0.81\pm0.21$} &\tiny{$0.85\pm0.22$} &\tiny{$0.88\pm0.23$} &\tiny{$0.91\pm0.23$ } \\
\hline$~f_{3}^{B_s^0\rightarrow {D_s^*}^+}(q^2)~$ &  \lower
0.30cm\hbox{{\vrule width 0pt height
1.cm}}\tiny{$-0.67\pm0.17$}&\tiny{$-0.69\pm0.17$}&\tiny{$-0.72\pm0.18$} &\tiny{$-1.29\pm0.32$}
&\tiny{$-1.42\pm0.34$} &\tiny{$-1.49\pm0.35$}&\tiny{$-1.55\pm0.36$}
\\ \hline
\end{tabular}
\vskip 0.3 cm

\end{center}
\caption{The values of form factors $f_{0,1,2,3}$ at different
values of $q^2$.}
\end{table}

\begin{table}[tbp]
\begin{center}
\begin{tabular}{||cccc||}
\hline\hline \lower0.35cm \hbox{{\vrule width 0pt height 1.0cm }}
$B_q\rightarrow D_qP$&present work  &PQCD \cite{pqcd}&Exp \cite{Yao}\\
\hline\hline \lower0.35cm \hbox{{\vrule width 0pt height 1.0cm }}
$B^{\pm}\rightarrow\bar{D^0}{\pi}^{\pm}$&$(5.95\pm1.95)\times10^{-3}$&$5.11^{+2.95+0.43+0.15}_{-2.07-0.75-0.15}\times10^{-3}$ & $(4.92\pm0.20)\times10^{-3}$\\
\lower0.35cm \hbox{{\vrule width 0pt height 1.0cm }}
$B^{\pm}\rightarrow\bar{D^0}{K}^{\pm}$&$(4.31\pm1.52)\times10^{-4}$ &$4.00^{+2.35+0.63+0.12}_{-1.64-0.93-0.12}\times10^{-4}$ &$(4.08\pm0.24)\times10^{-4}$\\
\lower0.35cm \hbox{{\vrule width 0pt height 1.0cm }}
$B^{\pm}\rightarrow\bar{D^0}{D}^{\pm}$&$(3.44\pm1.22)\times10^{-4}$ &$-$ & $(4.80\pm1.00)\times10^{-4}$\\
\lower0.35cm \hbox{{\vrule width 0pt height 1.0cm }}
$B^{\pm}\rightarrow\bar{D^0}{D_s}^{\pm}$&$(2.03\pm0.85)\times10^{-2}$ &$-$ & $(1.09\pm0.27)~\%$\\
\lower0.35cm \hbox{{\vrule width 0pt height 1.0cm }}
$B^{0}\rightarrow\bar{D}^{\pm}{\pi}^{\mp}$&$(5.69\pm1.70)\times10^{-3}$ &$2.69^{+1.78+0.55+0.08}_{-1.17-0.73-0.08}\times10^{-3}$ & $(3.40\pm0.90)\times10^{-3}$\\
\lower0.35cm \hbox{{\vrule width 0pt height 1.0cm }}
$B^{0}\rightarrow\bar{D}^{\pm}{K}^{\mp}$&$(3.53\pm1.23)\times10^{-4}$ &$2.43^{1.56+0.63+0.07}_{-1.01-0.71-0.07}\times10^{-4}$ & $(2.00\pm0.60)\times10^{-4}$\\
\lower0.35cm \hbox{{\vrule width 0pt height 1.0cm }}
$B^{0}\rightarrow\bar{D}^{\pm}{D}^{\mp}$&$(2.87\pm0.89)\times10^{-4}$ &$-$ & $(1.90\pm0.60)\times10^{-4}$\\
\lower0.35cm \hbox{{\vrule width 0pt height 1.0cm }}
$B^{0}\rightarrow\bar{D}^{\pm}{D_s}^{\mp}$&$(8.88\pm2.82)\times10^{-3}$ &$-$ & $(6.50\pm2.10)\times10^{-3}$\\
\lower0.35cm \hbox{{\vrule width 0pt height 1.0cm }}
$B_s^{0}\rightarrow{D_s}^{\pm}{\pi}^{\mp}$&$(1.42\pm0.57)\times10^{-3}$ &$2.13^{+1.14+0.69+0.06}_{-0.81-0.68-0.06}\times10^{-3}$ & $  (3.80\pm0.30)\times10^{-3}  $ \\
\lower0.35cm \hbox{{\vrule width 0pt height 1.0cm }}
$B_s^{0}\rightarrow{D_s}^{\pm}{K}^{\mp}$&$(1.03\pm0.51)\times10^{-4}$ &$1.71^{+0.92+0.58+0.05}_{-0.65-0.55-0.05}\times10^{-4}$ & $  - $\\
\lower0.35cm \hbox{{\vrule width 0pt height 1.0cm }}
$B_s^{0}\rightarrow{D_s}^{\pm}{D}^{\mp}$&$(1.20\pm0.73)\times10^{-4}$ &$-$ & $-$\\
\lower0.35cm \hbox{{\vrule width 0pt height 1.0cm }}
$B_s^{0}\rightarrow{D_s}^{\pm}{D_s}^{\mp}$&$(2.17\pm0.82)\times10^{-3}$&$-$&$-$\\
\hline\hline
\end{tabular}
\vskip0.3 cm \vskip1 cm

\end{center}
\caption{Values for the branching ratio of $B_q\rightarrow D_qP$.}
\end{table}

\begin{table}[tbp]
\begin{center}
\begin{tabular}{||cccc||}
\hline\hline \lower0.35cm \hbox{{\vrule width 0pt height 1.0cm }}
$B\rightarrow D_{q}^{*}P$& present work  & PQCD \cite{pqcd}& Exp \cite{Yao} \\
\hline\hline \lower0.35cm \hbox{{\vrule width 0pt height 1.0cm }}
$B^{\pm}\rightarrow\bar{D^*}^{0}{\pi}^{\pm}$&$(4.89\pm1.52)\times10^{-3}$ &$5.04^{+2.92+0.44+0.15}_{-2.04-0.73-0.15}\times10^{-3}$& $(4.60\pm0.40)\times10^{-3}$\\
\lower0.35cm \hbox{{\vrule width 0pt height 1.0cm }}
$B^{\pm}\rightarrow\bar{D^*}^{0}{K}^{\pm}$&$(3.38\pm1.04)\times10^{-4}$ &$3.60^{+2.33+0.62+0.12}_{-1.62-0.92-0.12}\times10^{-4}$& $(3.70\pm0.40)\times10^{-4}$\\
\lower0.35cm \hbox{{\vrule width 0pt height 1.0cm }}
$B^{\pm}\rightarrow\bar{D^*}^{0}{D}^{\pm}$&$(2.57\pm0.88)\times10^{-4}$ &$-$& $-$\\
\lower0.35cm \hbox{{\vrule width 0pt height 1.0cm }}
$B^{\pm}\rightarrow\bar{D^*}^{0}{D_s}^{\pm}$&$(11.03\pm2.91)\times10^{-3}$ &$-$& $(10.00\pm4.00)\times10^{-3}$\\
\lower0.35cm \hbox{{\vrule width 0pt height 1.0cm }}
$B^{0}\rightarrow\bar{D^*}^{\pm}{\pi}^{\mp}$&$(3.45\pm1.75)\times10^{-3}$ &$2.60^{+1.73+0.53+0.07}_{-1.14-0.70-0.07}\times10^{-3}$& $(2.76\pm0.21)\times10^{-3}$\\
\lower0.35cm \hbox{{\vrule width 0pt height 1.0cm }}
$B^{0}\rightarrow\bar{D^*}^{\pm}{K}^{\mp}$&$(2.08\pm0.68)\times10^{-4}$ &$2.37^{+1.52+0.62+0.07}_{-0.99-0.69-0.07}\times10^{-4}$& $(2.14\pm0.20)\times10^{-4}$\\
\lower0.35cm \hbox{{\vrule width 0pt height 1.0cm }}
$B^{0}\rightarrow\bar{D^*}^{\pm}{D}^{\mp}$&$(3.14\pm1.46)\times10^{-4}$ &$-$& $(9.30\pm1.50)\times10^{-4}$\\
\lower0.35cm \hbox{{\vrule width 0pt height 1.0cm }}
$B^{0}\rightarrow\bar{D^*}^{\pm}{D_s}^{\mp}$&$(8.69\pm2.88)\times10^{-3}$ &$-$& $(8.80\pm1.60)\times10^{-3}$\\
\lower0.35cm \hbox{{\vrule width 0pt height 1.0cm }}
$B_s^{0}\rightarrow{D_s^*}^{\pm}{\pi}^{\mp}$&$(2.11\pm0.73)\times10^{-3}$ &$2.42^{+1.12+0.78+0.07}_{-0.72-0.77-0.07}\times10^{-3}$& $ -  $ \\
\lower0.35cm \hbox{{\vrule width 0pt height 1.0cm }}
$B_s^{0}\rightarrow{D_s^*}^{\pm}{K}^{\mp}$&$(1.59\pm0.67)\times10^{-4}$ &$1.65^{+0.90+0.56+0.05}_{-0.63-0.53-0.05}\times10^{-4}$& $  - $\\
\lower0.35cm \hbox{{\vrule width 0pt height 1.0cm }}
$B_s^{0}\rightarrow{D_s^*}^{\pm}{D}^{\mp}$&$(0.30\pm0.11)\times10^{-4}$ &$-$&$-$\\
\lower0.35cm \hbox{{\vrule width 0pt height 1.0cm }}
$B_s^{0}\rightarrow{D_s^*}^{\pm}{D_s}^{\mp}$&$(2.54\pm0.57)\times10^{-3}$&$-$ &$-$\\
\hline\hline
\end{tabular}
\vskip0.3 cm \vskip1 cm

\end{center}
\caption{Values for the branching ratio of $B_q\rightarrow
D_q^{*}P$.}
\end{table}

\begin{table}[tbp]
\begin{center}
\begin{tabular}{||cccc||}
\hline\hline \lower0.35cm \hbox{{\vrule width 0pt height 1.0cm }}
$B_q\rightarrow D_qV$&present work  & PQCD \cite{pqcd}& Exp \cite{Yao}\\
\hline\hline \lower0.35cm \hbox{{\vrule width 0pt height 1.0cm }}
$B^{\pm}\rightarrow\bar{D}^{0}{K^*(892)}^{\pm}$&$(2.90\pm0.88)\times10^{-4}$ &$6.49^{+3.86+0.12+0.20}_{-2.68-1.58-0.20}\times10^{-4}$& $(6.30\pm0.80)\times10^{-4}$\\
\lower0.35cm \hbox{{\vrule width 0pt height 1.0cm }}
$B^{\pm}\rightarrow\bar{D}^{0}{D^*}^{\pm}$&$(5.61\pm1.88)\times10^{-4}$ &$-$&$(4.60\pm0.90)\times10^{-4}$\\
\lower0.35cm \hbox{{\vrule width 0pt height 1.0cm }}
$B^{\pm}\rightarrow\bar{D}^{0}{D^*_s}^{\pm}$&$(7.01\pm2.09)\times10^{-3}$ &$-$& $(7.20\pm2.60)\times10^{-3}$\\
\lower0.35cm \hbox{{\vrule width 0pt height 1.0cm }}
$B^{0}\rightarrow\bar{D}^{\pm}{K^*(892)}^{\mp}$&$(3.20\pm1.15)\times10^{-4}$ &$4.07^{+2.61+0.94+0.12}_{-1.69-1.11-0.12}\times10^{-4}$& $(4.50\pm0.70)\times10^{-4}$\\
\lower0.35cm \hbox{{\vrule width 0pt height 1.0cm }}
$B^{0}\rightarrow\bar{D}^{\pm}{D^*}^{\mp}$&$(8.18\pm2.84)\times10^{-4}$ &$-$& $-$\\
\lower0.35cm \hbox{{\vrule width 0pt height 1.0cm }}
$B^{0}\rightarrow\bar{D}^{\pm}{D_s^*}^{\mp}$&$(9.23\pm2.67)\times10^{-3}$&$-$&  $(8.60\pm3.40)\times10^{-3}$\\
\lower0.35cm \hbox{{\vrule width 0pt height 1.0cm }}
$B_s^{0}\rightarrow{D_s}^{\pm}{K^{*}(892)}^{\mp}$&$(0.50\pm0.22)\times10^{-4}$ &$3.02^{+1.62+0.88+0.10}_{-1.16-0.91-0.10}\times10^{-4}$& $  - $\\
\lower0.35cm \hbox{{\vrule width 0pt height 1.0cm }}
$B_s^{0}\rightarrow{D_s}^{\pm}{D^{*}}^{\mp}$&$(1.07\pm0.59)\times10^{-4}$ &$-$& $-$\\
\lower0.35cm \hbox{{\vrule width 0pt height 1.0cm }}
$B_s^{0}\rightarrow{D_s}^{\pm}{D_s^*}^{\mp}$&$(2.62\pm0.93)\times10^{-3}$ &$-$&$-$\\
\hline\hline
\end{tabular}
\vskip0.3 cm \vskip1 cm

\end{center}
\caption{Values for the branching ratio of $B_q\rightarrow D_qV$.}
\end{table}

\begin{table}[tbp]
\begin{center}
\begin{tabular}{||cccc||}
\hline\hline \lower0.35cm \hbox{{\vrule width 0pt height 1.0cm }}
$B_q\rightarrow D_q^{*}V$&present work & PQCD\cite{pqcd} &Exp\cite{Yao} \\
\hline\hline \lower0.35cm \hbox{{\vrule width 0pt height 1.0cm }}
$B^{\pm}\rightarrow\bar{D^*}^{0}{K^*(892)}^{\pm}$&$(5.07\pm2.61)\times10^{-4}$ &$6.82^{+4.14+1.22+0.21}_{-2.80-1.65-0.21}\times10^{-4}$& $(8.30\pm1.50)\times10^{-4}$\\
\lower0.35cm \hbox{{\vrule width 0pt height 1.0cm }}
$B^{\pm}\rightarrow\bar{D^*}^{0}{D^*}^{\pm}$&$(0.11\pm0.07)\times10^{-2}$ &$-$& $<1.1~\%$\\
\lower0.35cm \hbox{{\vrule width 0pt height 1.0cm }}
$B^{\pm}\rightarrow\bar{D^*}^{0}{D^*_s}^{\pm}$&$(6.85\pm2.98)\times10^{-2}$ &$-$& $2.20\pm0.70~\%$\\
\lower0.35cm \hbox{{\vrule width 0pt height 1.0cm }}
$B^{0}\rightarrow\bar{D^*}^{\pm}{K^*(890)}^{\mp}$&$(3.55\pm1.25)\times10^{-4}$ &$4.88^{+3.18+1.16+0.15}_{-2.08-1.41-0.15}\times10^{-4}$& $(3.30\pm0.60)\times10^{-4}$\\
\lower0.35cm \hbox{{\vrule width 0pt height 1.0cm }}
$B^{0}\rightarrow\bar{D^*}^{\pm}{D^*}^{\mp}$&$(8.78\pm2.50)\times10^{-4}$ &$-$& $(8.30\pm1.01)\times10^{-4}$\\
\lower0.35cm \hbox{{\vrule width 0pt height 1.0cm }}
$B^{0}\rightarrow\bar{D^*}^{\pm}{D_s^*}^{\mp}$&$(8.17\pm2.93)\times10^{-2}$ &$-$& $(1.79\pm0.16)~\%$\\
\lower0.35cm \hbox{{\vrule width 0pt height 1.0cm }}
$B_s^{0}\rightarrow{D^*_s}^{\pm}{K^{*}(890)}^{\mp}$&$(1.63\pm0.86)\times10^{-4}$ &$3.47^{+1.96+1.07+0.11}_{-1.35-1.66-0.11}\times10^{-4}$& $  - $\\
\lower0.35cm \hbox{{\vrule width 0pt height 1.0cm }}
$B_s^{0}\rightarrow{D^*_s}^{\pm}{D^{*}}^{\mp}$&$(6.76\pm2.69)\times10^{-4}$ &$-$& $-$\\
\lower0.35cm \hbox{{\vrule width 0pt height 1.0cm }}
$B_s^{0}\rightarrow{D^*_s}^{\pm}{D_s^*}^{\mp}$&$(2.77\pm0.76)\times10^{-2}$&$-$ &$(23^{+21}_{-13})~\%$\\
\hline\hline
\end{tabular}
\vskip0.3 cm \vskip1 cm

\end{center}
\caption{Values for the branching ratio of $B_q\rightarrow
D_q^{*}V$.}
\end{table}

\end{document}